\documentclass[aps,prl,twocolumn,superscriptaddress,showpacs]{revtex4-1}
\usepackage{graphicx,amsmath,amssymb,bm}

\newcommand{\be}{\begin{equation}}
\newcommand{\ee}{\end{equation}}
\newcommand{\mev}{\, \text{MeV}}
\newcommand{\pp}{{\bf p}}
\newcommand{\PP}{{\bf P}}
\renewcommand{\vec}[1]{\boldsymbol{#1}}

\renewcommand{\Im}{\textrm{Im}}

\renewcommand{\t}[1]{\tilde{#1}}
\newcommand{\mel}[3]{\left<#1\left|\vphantom{#1#3}#2\right|#3\right>}
\newcommand{\gccm}{\mbox{g\,cm}^{-3}}

\begin{document}

\title{Supernova matter at subnuclear densities as a resonant Fermi gas: \\
Enhancement of neutrino rates}

\author{A.\ Bartl}
\affiliation{Institut f\"ur Kernphysik, 
Technische Universit\"at Darmstadt, D-64289 Darmstadt, Germany}
\affiliation{ExtreMe Matter Institute EMMI, 
GSI Helmholtzzentrum f\"ur Schwerionenforschung GmbH, 
D-64291 Darmstadt, Germany}
\author{C.\ J.\ Pethick}
\affiliation{The Niels Bohr International Academy, 
The Niels Bohr Institute, University of Copenhagen, Blegdamsvej 17, 
DK-2100 Copenhagen \O, Denmark}
\affiliation{NORDITA, KTH Royal Institute of Technology and 
Stockholm University, Roslagstullsbacken 23, SE-106\;91 Stockholm, Sweden}
\author{A.\ Schwenk}
\affiliation{ExtreMe Matter Institute EMMI, 
GSI Helmholtzzentrum f\"ur Schwerionenforschung GmbH, 
D-64291 Darmstadt, Germany}
\affiliation{Institut f\"ur Kernphysik, 
Technische Universit\"at Darmstadt, D-64289 Darmstadt, Germany}

\begin{abstract}
At low energies nucleon-nucleon interactions are resonant and
therefore supernova matter at subnuclear densities has many
similarities to atomic gases with interactions dominated by a Feshbach
resonance. We calculate the rates of neutrino processes involving
nucleon-nucleon collisions and show that these are enhanced in
mixtures of neutrons and protons at subnuclear densities due to the
large scattering lengths. As a result, the rate for neutrino pair
bremsstrahlung and absorption is significantly larger below
$10^{13}$\,g\,cm$^{-3}$ compared to rates used in supernova
simulations.
\end{abstract}


\maketitle

{\it Introduction.--} In the standard model of core-collapse
supernovae, energy is transferred to outer parts of the star by
neutrinos diffusing out from the stellar core, thereby expelling
matter. Matter in the density range $\rho \sim 10^{11} -
10^{13}$\,g\,cm$^{-3}$ plays an important role after core bounce,
because under these conditions neutrinos decouple from the
matter~\cite{janka}. In this Letter, we calculate rates of neutrino
processes in this regime. We find that nucleon-nucleon (NN)
interactions, which are resonant at low energies (as is reflected in
the weak binding of the deuteron) significantly enhance neutrino
rates.

In neutral-current neutrino-nucleon processes, the axial vector part
of weak interactions dominates.  For a system consisting of neutrons
alone, only noncentral parts of NN interactions contribute, whereas
when both neutrons and protons are present, the central part also
enters because the axial charges of the neutron and proton are
unequal. In numerical simulations of stellar collapse, rates of
neutrino processes are commonly treated in the one-pion-exchange
approximation for NN interactions~\cite{hannestad98} and rates for
matter containing both neutrons and protons are then obtained by
replacing the neutron density by the total nucleon density. Neutrino
processes in degenerate neutron matter have been studied based on more
realistic NN scattering amplitudes by Hanhart {\it et al.}~\cite{hanhart01}, 
who used free-space scattering amplitudes expressed in terms of 
experimentally determined phase shifts, and by Bacca {\it et
al.}~\cite{bacca09,bacca12}, who also considered nondegenerate
conditions and chiral effective field theory (EFT)
interactions. Effects of neutron-proton collisions have been discussed
in a number of works, including that of Friman and
Maxwell~\cite{friman79} for degenerate matter in the context of
neutron star cooling, and that of Sigl~\cite{sigl97} directed towards
processes in supernovae.

In this Letter, we investigate the rate of neutrino processes in
matter at subnuclear densities taking into account the resonant nature
of NN interactions. The basic input for calculations of rates of
neutrino processes is the axial charge-density structure 
factor~\cite{raffelt_book}
\be
S_A({\bf q}, \omega) = \frac{\sum\limits_{m,n}  e^{-E_m/T} \bigl|
\langle m | \rho_A({\bf q}) | n \rangle \bigr|^2 
\delta(\omega-E_n+E_m) }{\sum\limits_m e^{-E_m/T}} \,,
\ee
where ${\bf q}$ and $\omega$ are the momentum and energy transfers, the 
states $|m \rangle$ and $|n \rangle$ are eigenstates of the nucleonic
system, $\rho_A({\bf q}) $ is the Fourier transform of the axial
charge-density operator, and $T$ is the temperature. $S_A$ is related
to the axial charge-density correlation function $\chi_A$ by
\be
S_A({\bf q}, \omega)=\frac1{\pi n} \frac1{1-e^{-\omega/T}} \,
\Im \, \chi_A({\bf q}, \omega) \,,
\ee
where $n$ is the total density of nucleons. (We work in units with
$\hbar=c=k_B=1$.)  From $S_A({\bf
q}, \omega)$ one can calculate the rates of neutrino scattering and
of neutrino pair creation and annihilation.

\begin{figure*}[t]
\begin{center}
\includegraphics[width=\textwidth]{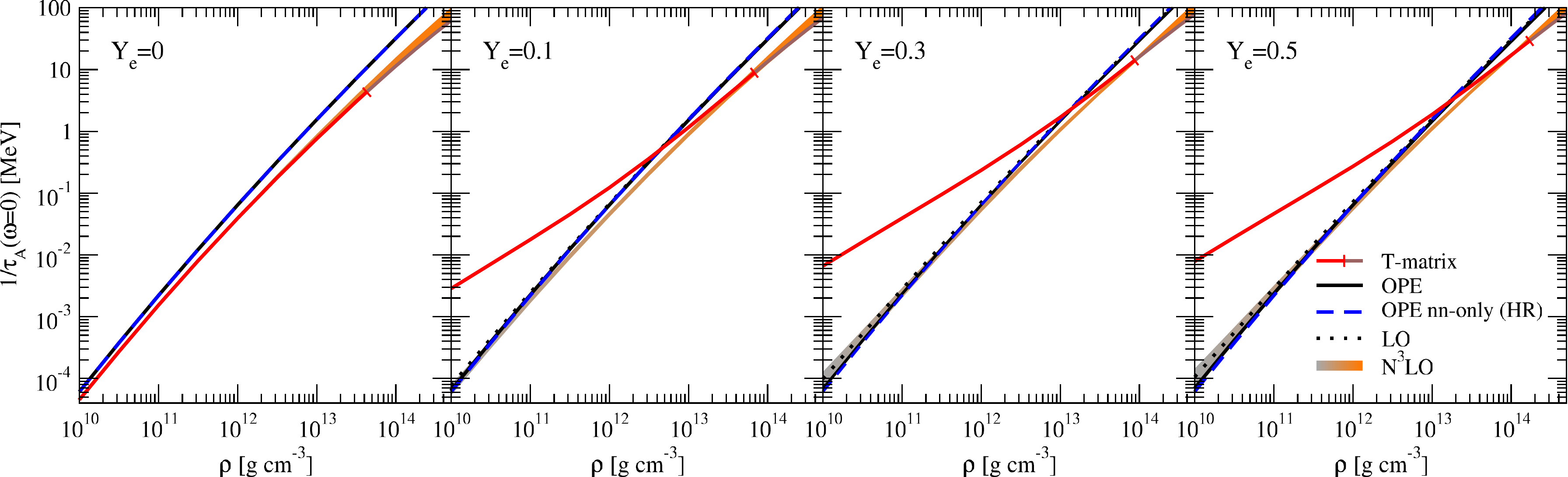}
\end{center}
\vspace{-4mm}
\caption{(Color online) Spin relaxation rate $1/\tau_A$ for $\omega=0$
as a function of density $\rho$ for different electron fractions $Y_e$. The temperature is
taken along typical supernova conditions,
Eq.~(\ref{SNconditions}). Results are shown for the OPE approximation,
the approximation used in supernova simulations
(OPE nn-only, HR)~\cite{hannestad98}, leading-order (LO) chiral EFT
interactions, and including NN interactions at N$^3$LO at the Born
level. In addition, we show the results based
on NN phase shifts ($T$-matrix). The color change from grey to orange
indicates that the N$^3$LO results should only be trusted at higher
densities where the Born approximation works well. The $T$-matrix
results are expected to be valid up to a fugacity of $z=1/2$, which is
marked by the small bar.\label{fig:tauA}}
\vspace*{-2mm}
\end{figure*}

In most calculations, the approach adopted is to use the
nonrelativistic limit for the coupling of the axial field to the
nucleons, in which case its strength is given by the spin operator
times the axial charge of the nucleon $C_A$. For small momentum and
energy transfers, the axial charge may be taken to be a constant, but
more generally there are momentum-dependent and two-body current
contributions~\cite{javier}. Strong interactions are included at different levels in
the structure factor: at low densities directly from two-body
scattering data, and in general based on NN interactions. One
technical point is that the scattering amplitudes required to
calculate the structure factor are generally off-shell ones, and
therefore it is necessary to specify the energy at which the
scattering amplitude is evaluated. This is particularly important at
low energies because the interactions are resonant. The aim of the
present work is to give a first estimate of the effects of NN
scattering, and therefore we shall, as a first approximation, evaluate
scattering amplitudes for an energy equal to the mean of the energies
of the initial and final states. A deeper investigation of this
problem is left for future work (see also Ref.~\cite{Liou} in the
context of $pp \to pp \gamma$). There are also contributions from
three- and higher-body interactions, but at subnuclear densities,
which appear to be particularly important in core-collapse supernovae,
these are expected to be small.

{\it Axial response function.--} The basic picture that we shall adopt
is to consider the nuclear medium as a system of interacting
quasiparticles~\cite{bacca09,bacca12,lykasov_et_al05,lykasov_et_al08}. To
make the treatment more transparent, we do not include mean-field
effects, because at subnuclear densities they are expected to be
relatively unimportant for neutral currents. We consider the 
long-wavelength limit ($\vec q \to 0$), which is a good approximation
for bremsstrahlung processes. For inelastic scattering, recoil effects
need to be included. The generalization of our formalism to finite
$\vec q$, and the interesting interplay of the widths ($1/\tau_A$)
with recoil effects will be studied in a future paper. For zero
frequency, $\chi_A$ at long wavelengths is given by~\cite{burrows98}
\be
\chi^0_A = \sum_{i=np} \sum_1 (C_A^i\sigma_1)^2
\left( -\frac{\partial n_1^i}{\partial \epsilon_1^i} \right) \,.
\ee
Here we use the shorthand notation $1 \equiv ({\bf p}_1,\sigma_1)$ for
momentum and spin. $n_1^i$ is the quasiparticle distribution function
for species $i=n,p$, and $\epsilon_1^i$ is the quasiparticle
energy. Our approach to determine the frequency dependence follows
Refs.~\cite{hannestad98,sigl97,raffelt_book} by
calculating the response at frequencies high compared with a typical
collision frequency, and then extending the results to low 
frequency by assuming that the response function has the standard
(Debye) form for a simple relaxation process.

To calculate $\chi_A$ at frequencies high compared with typical
collision rates for nucleons, we generalize the formalism of
Ref.~\cite{bacca12} to mixtures of neutrons and protons. The only
difference compared with the result for pure neutrons is that the
particles can be either neutrons or protons. For $q=0$, we find
\be
\chi_A \approx \frac{\rm i}{\omega} \frac{\chi_A^0}{\tau_A} \equiv
\frac{\rm i}{\omega} \Upsilon_A \hspace{2mm} {\rm with} \hspace{2mm}
\Upsilon_A = \Upsilon_A^{nn}+\Upsilon_A^{pp}+2\Upsilon_A^{np} \,,
\label{upsA}
\ee
where the superscripts indicate the nucleons involved in the process.
Finally, the expression for $\chi_A$ that interpolates between the
low- and high-frequency limits is
\be
\chi_A = \frac{\chi_A^0}{1 - {\rm i} \omega \tau_A} \,.
\ee
Support for our ansatz comes from exact solutions of the Boltzmann 
equation for both degenerate and classical gases (see Ref.~\cite{PS}
and references therein), which show that, for those conditions, the
effective relaxation time for $|\omega \tau_A| \ll 1$ differs from that
for $|\omega \tau_A | \gg 1$ by less than $10\%$. As a preview of our
results for the spin relaxation rate in mixtures of neutrons and
protons, we refer to Fig.~\ref{fig:tauA}.

The different NN contributions $\Upsilon_A^{ij}$ are given by
(for details on the formalism, we refer the reader to 
Ref.~\cite{bacca12})
\begin{align} 
\Upsilon_A^{ij} &= \frac{1}{1+\delta_{ij}} \frac{2\pi}{4\omega} \sum_{1234} 
\delta^+_\epsilon \delta_{\bf p} \left| \mel{34}{{\cal T}^{ij}}{12}\right|^2 
\nonumber \\
&\times\left[n^i_1 n^j_2 (1-n^i_3)(1-n^j_4)
-n^i_3 n^j_4 (1-n^i_1)(1-n^j_2) \right] \nonumber \\
&\times (C_A^i\sigma_1+C_A^j\sigma_2-C_A^i\sigma_3-C_A^j\sigma_4)^2 \,,
\label{upsilonij}
\end{align}
which is an even function of $\omega$. Here, $\delta^+_\epsilon \equiv
\delta(\omega + \epsilon^i_1+\epsilon^j_2-\epsilon^i_3-\epsilon^j_4)$
and $\delta_{\bf p} \equiv \delta ( {\bf p}_1 +{\bf p}_2-{\bf
p}_3-{\bf p}_4)$ are energy and momentum conserving delta
functions. We will employ scattering amplitudes ${\cal T}^{ij}$ that
include exchange contributions, and the factor of $1/(1+\delta_{ij})$ in
Eq.~(\ref{upsilonij}) is to avoid double-counting of final states
for collisions between particles of the same species. Thus
\begin{align}
&\Upsilon_A^{ij} = \frac{\pi}{\omega} \sum_{1234} 
\delta^+_\epsilon \delta_{\bf p} \left| \mel{34}{{\cal T}^{ij}}{12}\right|^2
\nonumber \\
&\times\left[n^i_1 n^j_2 (1-n^i_3)(1-n^j_4)
-n^i_3 n^j_4 (1-n^i_1)(1-n^j_2) \right] \nonumber \\
&\times
\begin{cases}
C_A^{i\,2} \sigma_1 (\sigma_1+\sigma_2-\sigma_3-\sigma_4)  
& \mbox{for } i=j, \\[2mm]
C_A^{i\,2} \sigma_1 (\sigma_1-\sigma_3)+ C_A^{j\,2}\sigma_2 (\sigma_2-\sigma_4)\\
\quad +2 C_A^iC_A^j\sigma_1 (\sigma_2-\sigma_4)
& \mbox{for } i\neq j.
\end{cases}
\end{align}

To separate the spin sums from the phase space integration in 
$\Upsilon_A^{ij}$, we introduce the quantities 
\begin{align}
&W^{ij} = \frac1{12} \sum_{\sigma_i} \Big[ \langle 34 |{\cal T}^{ij}|12 
\rangle^* \, \boldsymbol\sigma_1 \nonumber \\
&\times
\begin{cases}
C_A^{i\,2} \left[\boldsymbol\sigma_1+\boldsymbol\sigma_2,
\langle 34 |{\cal T}^{ij}|12 \rangle\right] \Big] & \mbox{for }i=j, \\[2mm]
(C_A^{i\,2}+C_A^{j\,2}) \left[\boldsymbol\sigma_1,
\langle 34 |{\cal T}^{ij}|12 \rangle\right] \\
\quad +2\,C_A^iC_A^j \left[\boldsymbol\sigma_2,
\langle 34 |{\cal T}^{ij}|12 \rangle\right]\Big] & \mbox{for } i\neq j.
\end{cases}
\label{eq:spintraces}
\end{align}
The $W^{ij}$ depend only on $P, p, p'$ and three angles specifying the
orientation of the total momentum $\PP=\pp_1+\pp_2=\pp_3+\pp_4$ and the
relative momenta $\pp=(\pp_1-\pp_2)/2$ and $\pp'=(\pp_3-\pp_4)/2$.  As
for pure neutrons~\cite{bacca12}, we can write the expressions for
$\Upsilon_A^{ij}$ in alternative forms by using
$n^i_\lambda/(1-n^i_\lambda)=e^{-(\epsilon^i_\lambda-\mu^i)/T}$ and
the invariance of interactions under time reversal. We shall assume
that the quasiparticle energy is of the form $\epsilon^i_\pp =
\epsilon^i_0 + p^2/(2m^*_i)$, where $m^*_i$ is the effective mass.

In the nondegenerate limit, $\chi^0_A=\sum_{i=np} C_A^{i\,2} \,
n_i/T$, and if ${\cal T}^{ij}$ is independent of $\PP$, we can finally
write
\begin{align}
\Upsilon_A^{ij} &= \frac{n_i n_j \, \sinh (\omega/2T)}{\omega \pi 
\sqrt{2\pi m_{ij}^* T^3}} \, e^{-\omega/2T} \int\limits_0^\infty 
dp \, p^2\, e^{-p^2/(2m_{ij}^*T)} \nonumber \\
&\quad \times \int\limits_{-1}^1 d\cos \theta \, \sqrt{p^2+2m_{ij}^*\omega}
\, W^{ij} \,,
\end{align}
where $m_{ij}^*$ is the reduced mass for quasiparticles of species $i$
and $j$, and $\cos \theta = \widehat{\PP} \cdot \widehat{\pp}$. For
definiteness, we have taken $\omega$ to be positive. For negative
$\omega$, the lower limit of the $p$ integral is determined by the
condition that the square root vanishes. 

In addition, we can expand $\Upsilon_A^{ij}$ in partial waves:
\begin{widetext}
\begin{align}
&\Upsilon_A^{ij} = 16\sqrt\pi \, n_in_j \frac{\sinh(\omega/2T)}{\omega
\sqrt{2m_{ij}^*T^3}} \int\limits_0^\infty dp \, p^2 
\sqrt{p^2+2m_{ij}^*\omega} \, e^{-p^2/(2m^*_{ij}T)-\omega/(2T)} 
\sum_{S\t ST\t T}\sum_{Lll'J\t J} (-1)^{L+J+\t J} 
\left(\hat L\hat J\hat{\t J} \, \right)^2
\frac{\hat l \hat l'}{\hat S\hat{\t S}} \nonumber \\
&\times \left(1-(-1)^{l+S+T\vphantom{\t T}}\right) 
\left(1-(-1)^{l+\t S+\t T}\right) 
\left\{\begin{array}{ccc}l'&l&L\\l&l'&0\end{array}\right\} 
\left\{\begin{array}{ccc}l'&l&L\\S&S&J\end{array}\right\} 
\left\{\begin{array}{ccc}l&l'&L\\\t S&\t S&\t J\end{array}\right\} 
\mel{p}{\mathcal{T}^{ij,T}_{l'lJS}}{p'} 
\mel{p'}{\mathcal{T}^{ij,\t T}_{ll'\t J\t S}}{p} \nonumber \\
&\times \hspace{-1.5mm} \sum_{M_SM_S'} \hspace{-1.5mm} 
C_{L\Delta M_SSM_S'}^{SM_S} C_{L\Delta M_S\t SM_S'}^{\t SM_S} 
\begin{cases}
C_A^{i\,2} \, M_S \, \Delta M_S & \mbox{for } i=j, \\[2mm]
\frac18 (C_A^{i\,2}+C_A^{j\,2})(1-M_SM_S') 
+ \frac12 C_A^iC_A^j\left(M_S\,\Delta M_S-\frac12 (1-M_SM_S')\right)
& \mbox{for } i\neq j.
\end{cases}
\end{align} 
\end{widetext}
where $\widehat{a} \equiv \sqrt{2a+1}$ and $\Delta M_S \equiv M_S-M_S'$, 
and the sums for $i=j$ collapse to $S=\t S=T=\t T=1$.

{\it Results.--} Because we focus on relatively low densities, we
neglect effective mass effects and use $m^*_n=m^*_p=939 \mev$. For the
axial charges, we take $C_A^p=-C_A^n=g_A/2=1.26/2$ and note that
strange quark contributions, as discussed in Ref.~\cite{raffelt95},
are rather uncertain and do not change our results significantly. We
consider different electron fractions $Y_e$ from pure neutron to
symmetric matter and take for the temperature $T$ typical values 
in supernovae~\cite{bacca12},
\begin{equation}
T = 3 \mev \left(\frac{\rho}{10^{11}\,\rm g\,cm^{-3}}\right)^{1/3} \,,
\label{SNconditions}
\end{equation}
which corresponds to nondegenerate matter, except for $\rho >
10^{14}$\,g\,cm$^{-3}$ where neutrino processes are ineffective.

Figure~\ref{fig:tauA} shows the spin relaxation rate $1/\tau_A$ for
$\omega=0$ as a function of $\rho$ for different $Y_e$. First, we
consider the one-pion-exchange (OPE) approximation~\cite{check}, as
well as the typical approximation used in supernova simulations (OPE
nn-only, HR), which uses the neutron-neutron OPE rates of Hannestad
and Raffelt~\cite{hannestad98} also for neutron-proton mixtures by
replacing the neutron density by the total nucleon density. Note
that we only apply this prescription for the results labeled
HR. Figure~\ref{fig:tauA} shows that the results at the OPE level are
largely insensitive to the proton fraction.

A qualitatively similar dependence is found including all NN
interactions at N$^3$LO (also at the Born level), where the band in
Fig.~\ref{fig:tauA} is spanned by the EM $500\mev$, EGM $450/500\mev$
and EGM $450/700\mev$ potentials~\cite{entem03,epelbaum05}. These
chiral EFT interactions were recently found to be perturbative at
nuclear densities in neutron matter~\cite{IT}. At the N$^3$LO level,
we find a very weak dependence on $Y_e$ as well. As in the case of
pure neutron matter~\cite{bacca12}, the N$^3$LO rates are typically a
factor of two smaller at higher densities than the OPE approximation,
while they are similar at lower densities.

\begin{figure}[t]
\begin{center}
\includegraphics[width=0.9\columnwidth]{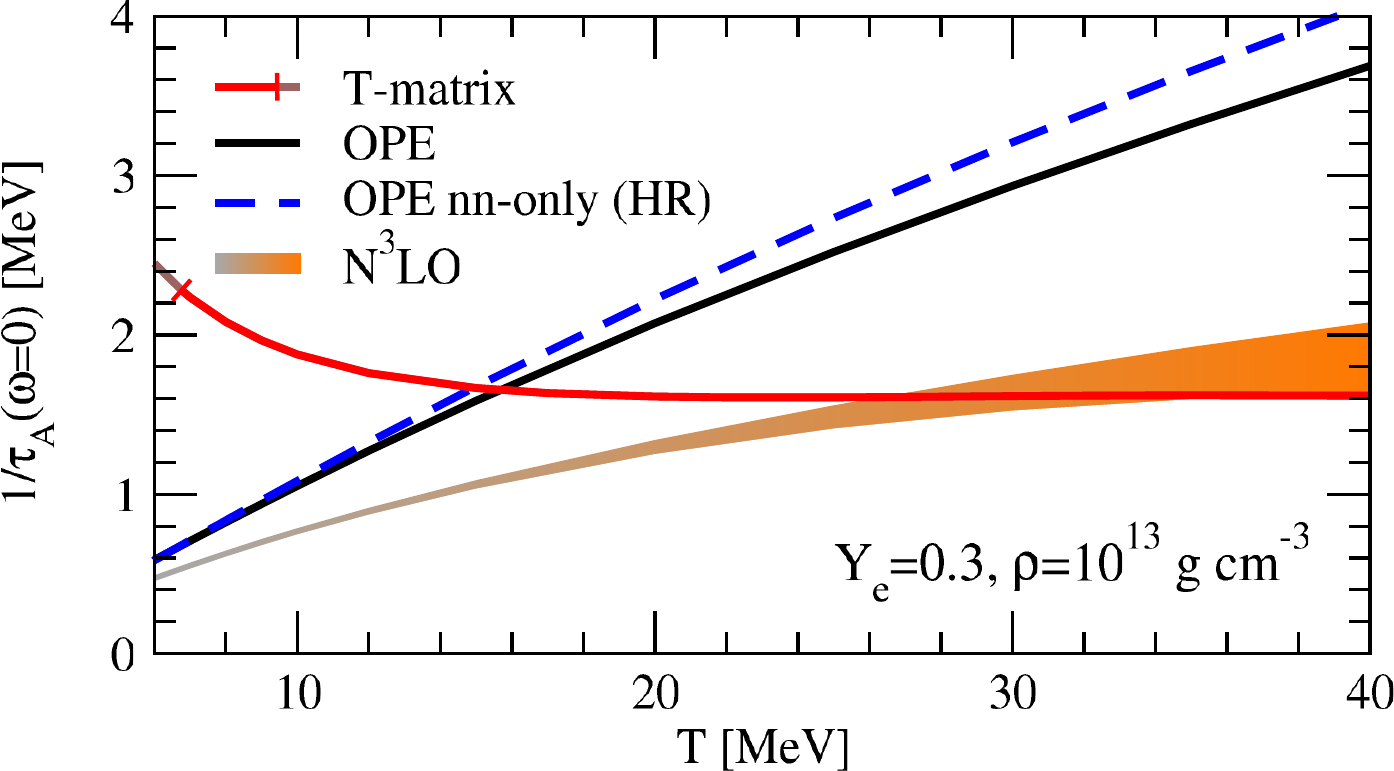}
\end{center}
\vspace{-2mm}
\caption{(Color online) Spin relaxation rate $1/\tau_A(\omega=0)$ as a
function of temperature $T$ for $Y_e=0.3$ and $\rho=10^{13}$\,g\,cm$^{-3}$.
Results are shown for the different cases as in
Fig.~\ref{fig:tauA}.\label{fig:tauA_constantrho}}
\end{figure}

At low densities, the typical momenta are also low, so that the
leading-order (LO) chiral EFT interactions are reliable. These
include, in addition to OPE, two central contact interactions
$V^{\rm LO}_{\rm ct} = C_S + C_T
\boldsymbol\sigma_1\cdot\boldsymbol\sigma_2$. For neutrons only,
$V^{\rm LO}_{\rm ct}$ does not contribute because it commutes with the
total spin $\boldsymbol\sigma_1+\boldsymbol\sigma_2$ in $W^{ii}$. However,
due to the different axial charges for neutrons and protons, the
spin-spin part $C_T$ contributes in mixtures. This leads to an
increase at low densities compared to OPE, shown in
Fig.~\ref{fig:tauA} by the dotted line (LO) for the EM $500\mev$ value
of $C_T$. This increase is small due to the approximate Wigner
symmetry with large scattering lengths in both S waves, implying a
small $C_T$ value.

\begin{figure}[t]
\begin{center}
\includegraphics[width=0.9\columnwidth]{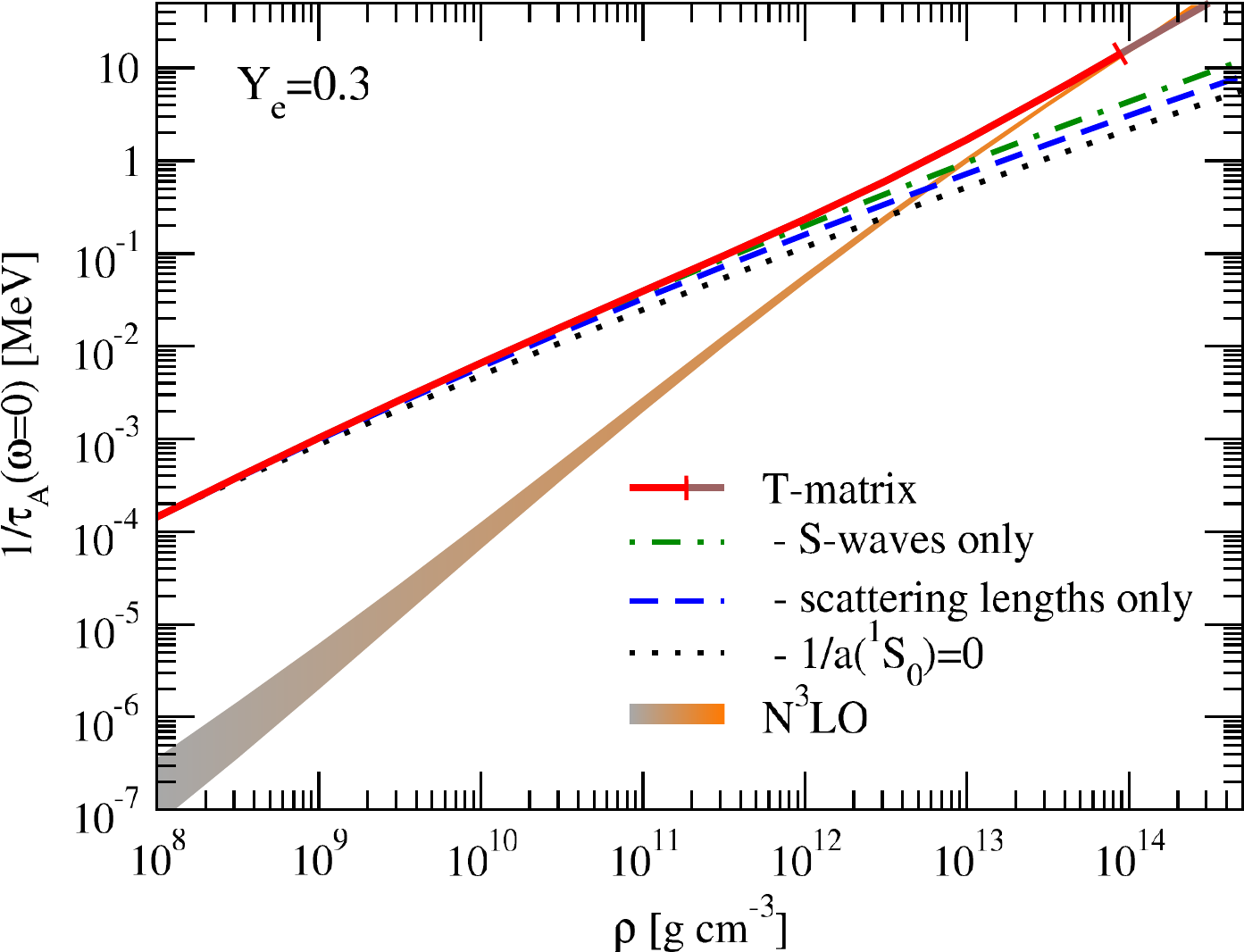}
\end{center}
\vspace{-2mm}
\caption{(Color online) $T$-matrix results for $1/\tau_A(\omega=0)$ as 
a function of density $\rho$ for $Y_e=0.3$, with temperature along
supernova conditions~(\ref{SNconditions}). The full $T$-matrix results
(solid line) are compared to only S-waves (dash-dotted
line), only S-wave scattering lengths (dashed line), and finally also
setting the $^1$S$_0$ scattering length to infinity (dotted line). For
comparison, the N$^3$LO results are also shown.\label{fig:tauA_tmatrix}}
\end{figure}

At low energies, NN interactions are resonant, so it is necessary to
go beyond the Born approximation. For low densities and nondegenerate
conditions, the spin relaxation rate can be determined model
independently from the $T$-matrix based on NN phase shifts, similar to
the virial expansion for energy contributions~\cite{virial}. The
resulting $1/\tau_A$ based on the Nijmegen partial wave
analysis~\cite{stoks93} is shown in Fig.~\ref{fig:tauA}. For neutron
matter, they agree well with the N$^3$LO results~\cite{bacca12},
because central interactions do not contribute. In mixtures of
neutrons and protons, we find a dramatic enhancement at subnuclear
densities $\rho \lesssim 10^{13}$\,g\,cm$^{-3}$ compared to the OPE
rates used in supernova simulations. This enhancement is due to the
large scattering lengths (see Fig.~\ref{fig:tauA_tmatrix}). The
$T$-matrix results are expected to be valid up to a fugacity of $z
\approx n_n\lambda_n^3/2 \lesssim 1/2$, where $\lambda_n$ is the
thermal wavelength. This is indicated by the small bar in
Fig.~\ref{fig:tauA}. Interestingly, around this point and for higher
densities, the $T$-matrix and N$^3$LO results agree well. This is
because NN interactions become weaker at higher energies. In general,
we expect higher-order $T$-matrix corrections to scale with a density
of states times a $T$-matrix. At low densities, for nondegenerate
conditions, and infinite scattering lengths, we thus expect
corrections to be of the order of $(n/T) 4 \pi/m \sqrt{3mT} \sim 0.02
- 0.07$ for $10^{11-12}$\,g\,cm$^{-3}$ and the supernova conditions
studied here.

In the nondegenerate limit, the energy scale of the collision
is set by the temperature. Therefore, we find the same
enhancement of the rate with decreasing temperature, as shown in
Fig.~\ref{fig:tauA_constantrho}. For higher temperatures, both
$T$-matrix and N$^3$LO results are a factor
of $\sim 2$ smaller compared to the OPE approximation.

The enhancement of the rates can be traced to the large scattering
lengths. To this end, we study in Fig.~\ref{fig:tauA_tmatrix} various
approximations for the $T$-matrix. At low densities, $1/\tau_A$ is
dominated by the S-wave contributions, mostly from the scattering
lengths. If we only keep the scattering lengths and also take
$1/a(^1\mbox{S}_0) = 0$, the low-density behavior can be reproduced with a
simple expression characterized by the $^3\mbox{S}_1$ scattering
length alone,
\be
\frac{1}{\tau_A(\omega=0)} \approx \frac{8\pi\,n_n n_p\,x}{n 
\sqrt{2\pi\,T}(m^*_{np})^{3/2}} \, e^x\,\Gamma(0,x) \,,
\ee
where $1/x=2m^*_{np}T\left(a(^3\mbox{S}_1)\right)^2$ and $\Gamma$
is the incomplete gamma function.

\begin{figure}[t]
\begin{center}
\includegraphics[width=0.9\columnwidth]{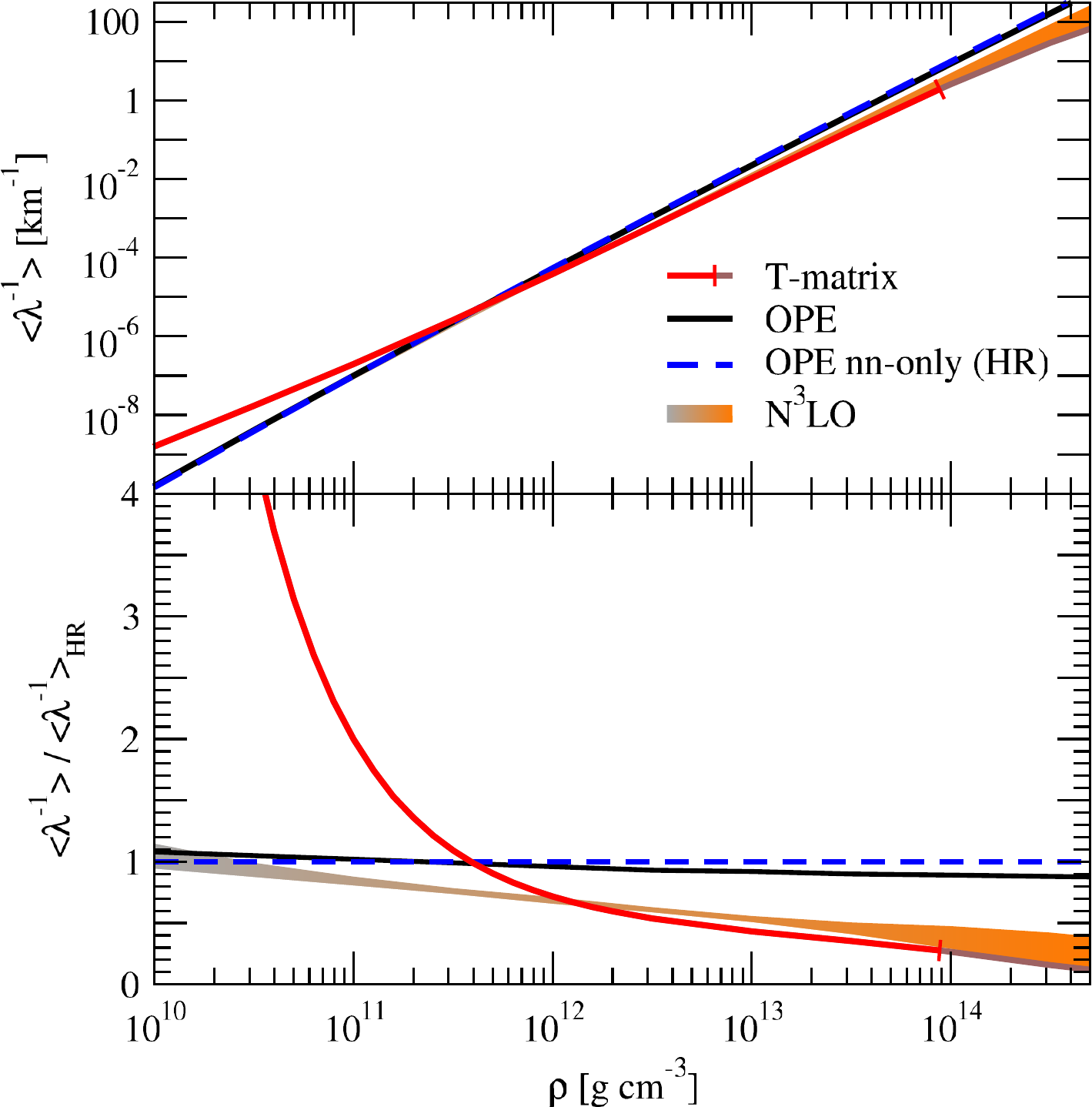}
\end{center}
\vspace{-2mm}
\caption{(Color online) Top panel: Energy-averaged inverse mean-free
path $\langle \lambda^{-1} \rangle$ of a neutrino against pair
absorption as a function of density $\rho$ for $Y_e=0.3$, with
temperature along supernova conditions~(\ref{SNconditions}). We assume
a Boltzmann distribution for the neutrino and antineutrino. Bottom
panel: Same, but normalized to the approximation used in supernova
simulations (OPE nn-only, HR)~\cite{hannestad98}.\label{fig:invmfp}}
\end{figure}

To explore the astrophysical impact of our findings, we show in
Fig.~\ref{fig:invmfp} the energy-averaged inverse mean-free path of a
neutrino against pair absorption (see
Refs.~\cite{hannestad98,lykasov_et_al08}). For the
conditions~(\ref{SNconditions}), the inverse mean-free path is
enhanced for $\rho \lesssim 10^{12}\,\gccm$. 
The enhancement is not as strong as in previous figures
because the inverse mean-free path contains an integral over
the energy exchange $\omega$ and the spin relaxation rate
based on the T-matrix formalism decreases faster with increasing
$\omega$ than the chiral EFT and OPE rates. For a fixed
neutrino energy, the opacity enhancement increases with decreasing
neutrino energy. Figure~\ref{fig:invmfp} shows again how close the
$T$-matrix and N$^3$LO results are for higher densities.  Combined
with the reduction of the opacity at higher densities, this can
contribute to energy transport from hotter matter at higher densities
to regions further out (see the comparison in the bottom panel to the
OPE approximation used in supernovae). This requires detailed
simulations that include competing neutrino processes at these
densities. Finally, the enhanced rates in mixtures also contribute to
inelastic scattering from NN pairs, which is the analog of neutrino
deuteron breakup when deuterons are dissolved~\cite{inelastic}.

In this Letter, we have studied neutrino processes involving NN
collisions in supernova matter at subnuclear densities. Due to the
resonant nature of NN interactions this regime has many similarities
to atomic gases with interactions dominated by a Feshbach
resonance. After generalizing the relaxation rate formalism to
mixtures of neutrons and protons, we have shown that in mixtures the
rates for neutrino pair bremsstrahlung and absorption are enhanced for
$\rho \lesssim 10^{13}$\,g\,cm$^{-3}$ due to the large scattering
lengths. Compared to rates used in supernova simulations, we find a
reduction of the rates at higher densities. Combined with the
enhancement at lower densities, this can provide an interesting
mechanism for energy transport to the outer layers.

We thank S.\ Bacca, A.\ Gezerlis, and H.-Th.\ Janka for useful
discussions. This work was supported by BMBF ARCHES, the ERC Grant
No.~307986 STRONGINT, the Helmholtz Alliance HA216/EMMI, and the
Studienstiftung des deutschen Volkes. This work was also supported
in part by NewCompStar, COST Action MP1304.


\begin{thebibliography}{99}
\bibitem{check} We have checked that the spin sums $W^{ij}$ agree
with previous results for neutron-proton bremsstrahlung of
Refs.~\cite{friman79,yakovlev01} (note the typo in~\cite{friman79},
see~\cite{yakovlev01}).
\end{thebibliography}
\end{document}